\documentclass{article}
\usepackage{spconf}
\usepackage{bm,amsmath,amssymb,color,graphicx}
\usepackage{algorithmic}
\usepackage[linesnumbered,ruled,vlined]{algorithm2e}
 
\title{Robust On-line Matrix Completion on Graphs}

\name{Symeon Chouvardas, Mohammed Amin Abdullah, Lucas Claude, Moez Draief}
 \address{Mathematical and Algorithmic Sciences Lab,\\
Huawei France R\&D,\\
Paris, France.\\
}

 \newcommand{\Real}{\mathbb{R} } 
  \newcommand{\tr}{\mathrm{tr} } 
    \newcommand{\vect}{\mathrm{vec} } 
   \newcommand{\Capr}{\widehat{C}_t }
   \newcommand{\qedsymbol}{$\blacksquare$}

\newtheorem{lem}{Lemma}
\newtheorem{thm}{Theorem}

\begin{document}
\maketitle
\begin{abstract}
We study online robust matrix completion on graphs. At each iteration a vector with some entries missing is revealed and our goal is 
to reconstruct it by identifying the underlying low-dimensional subspace from which the vectors are drawn. We assume there is an underlying graph structure to the data, that is, the components of each vector
correspond to nodes of a certain (known) graph, and their values are related accordingly. We give algorithms that exploit the graph to reconstruct the incomplete data, even in the presence of outlier noise. The theoretical properties of the algorithms are studied 
and numerical experiments using both synthetic and real world datasets verify the improved performance of the proposed technique compared to other state of the art algorithms. 
\end{abstract}

\section{INTRODUCTION}
The science of data acquisition, processing and inference has been boosted in recent years due to both the abundance of data and the economic benefits
associated with understanding it. Modern technologies have produced a vast array of data-generating devices (e.g., smart phones, cameras, sensors) and processes
(e.g., web searches, surveys, social media interaction). Frequently, it is known or conjectured that there is an underlying \emph{structure} to the data, and further, 
that this structure reflects some simpler underlying process. To make this more concrete, consider the notion of \emph{sparsity} in a movie-rating database such as used by
Netflix. If the rows of the ratings matrix represent movies and the columns people, then it is reasonable to suppose, and indeed has been found to be the case (\cite{netflix2,netflix1}), that the ratings 
matrix is of low rank. This would reflect that ratings vectors really ``live''  in a small subspace of the ambient space they are generated in. Consequently, understanding this subspace
better allows the exploitation of the data, for example, through more tightly focused marketing. On the other hand, the matrix will be incomplete, since any given 
user will rate only a very small subset of all the movies available. Thus, one would desire \emph{completion} of the matrix: a low-rank matrix with
sufficiently many observed entries can be \emph{exactly} reconstructed, and in the last decade, this has been a very active area of research in the signal processing and machine learning communities.
However, low-dimensional subspaces are not the form of structure; indeed, there can also be an underlying \emph{graphical} structure that represents connections/relations between entities. 
For example, one may generate a graph where movies are nodes and and two movies are linked if they both star a famous actor. Such a structure is sometimes easy to discover, and it immediately 
begs the question of how (if at all) it can be used to help fill in the missing entries of the matrix. 

The above is the subject of this paper: we investigate how graph structure can aid the reconstruction of a low rank matrix with missing entries, and specifically, in the \emph{on-line} setting. 
In this setting each column of the matrix (with some entries missing) is presented one at a time, and the algorithm must make the best estimation using only what has been presented so far. 
This is in contrast to the batch setting where the entire non-complete matrix is available to process. The motivations for the online setting are at least two-fold. Firstly, it more realistically 
reflects many situations. In the Netflix example, one may have a data stream of ratings. Secondly with the massive amount of data being generated, computational and memory limitations present
very real challenges to algorithms which operate in batch mode; indeed, it maybe impractical to even hold the entire matrix in memory, let alone perform complex operations on it.  

\vspace{-0.5cm}
\subsection{RELATED WORK}
The problem of matrix completion is a  well studied one and
several solutions have been proposed during the past years,
see for example \cite{CaRe09,CaTa10,candes2011robust}.
The online setup has its roots on the so--called subspace
tracking problem, e.g., 
\cite{PAST}, in which the columns of a matrix are revealed sequentially one per iteration step
and the goal is the identification of the underlying subspace. 
Extensions of these works, which deal with the presence
of missing entries and/or outliers have been studied in 
\cite{giannakis,rpca,Grasta,PETRELS,guo2014online,chouvardas2015robust}.
The batch version of the matrix completion on graphs problem
was originally presented in \cite{KaBrBrVa14} 
and extended to its robust version, which deals with the presence 
of outliers,  in 
\cite{shahid2015fast}.

\subsection{OUR CONTRIBUTION}
In this work, we extend the  idea  presented in 
\cite{KaBrBrVa14} and we propose a robust
online algorithm for matrix completion exploiting
graph information.
 Here, we propose an online solution,
 i.e., the columns of the matrix appear and are processed 
 sequentially, one per iteration step. To that direction,
 at each iteration step we define  a proper cost function
 and we minimize it to produce the updated estimates.
 Furthermore, we 
 study the case where there is 
 outlier noise, which corrupts a small subset of the observed vector.
We propose a robust solution, which estimates the outlier noise 
and cleans the data before updating the quantities of interest.
Our work has two notable differences compared to other online
matrix completion works. First, all other works do not 
exploit any graph information. Second, due to the absence
of the graph information, the problem they solve
decouples over the rows of the unknown subspace, which is not
the case here. This introduces a new difficulty.

\emph{Notation:}
Lowercase and uppercase boldfaced letters stand for
vectors and matrices respectively. The stage of discussion
will be $\mathbb{R}^{m\times r}$, where  the symbol
  $\mathbb{R}$ stands for the set of real numbers.
  Furthermore, $\Vert \bm{A} \Vert$ is the operator norm and $\Vert\bm{A}\Vert_F$ 
 the Frobenius norm of matrix $\bm{A}$.
   $\Vert\bm{x}\Vert, \ \Vert\bm{x}\Vert_1 $ denote
 the Euclidean and the $\ell_1$ norms of vector $\bm{x}$, respectively.
The symbol $\otimes$ stands for the Kronecker product.
Finally, $\bm{I}_{mr}$ is the $mr\times mr$ identity matrix
and $\bm{O}_{a\times b}$ is the zero matrix of dimension $a\times b$.
\section{MATRIX COMPLETION ON GRAPHS}
In this paper, we are concerned with the problem of matrix completion (MC) on graphs.
  The original task of  MC, e.g., \cite{CaRe09,Th15},
is the recovery of a data matrix from
a sample of its entries. Formally, given a matrix $\bm{X}$ of dimension
$m\times n$ we have access to $k \ll m\cdot n$ entries and the goal is the prediction
of the  rest unobserved ones. It has been shown that under
certain conditions  
this can be achieved   \cite{CaTa10,CaRe09}. 
Intuitively,
 MC builds upon the observation that if 
a certain matrix is structured, in the sense that it is of  low rank
or of approximate   low rank, then it can be recovered exactly, under
some mild assumptions regarding the  positions of the observed entries.  
The problem 
can be summarized as follows: Compute a  matrix,  $\bm{A}$, 
which will be of low rank and equal to the observation matrix $\bm{X}$ 
in the set of observed entries, say $\Omega$; that is
 $A_{ij}=X_{ij}, \forall i,j\in\Omega$,
where   $X_{ij}, \ A_{ij}$ is the
$i,j$--th entry of $\bm{X}$ and $\bm{A}$ respectively. 
 A way to do so is to solve the following problem: 
\begin{align*}
\min_{\bm{A}} \ &\mathrm{rank}(\bm{A}) \\ 
 s.t. \,\, & A_{ij} = X_{ij}, \ \forall i,j \in \Omega.
\end{align*}
The rank minimization problem described previously cannot be solved efficiently, since it is NP-hard \cite{CaRe09}. However,
it has been shown, \cite{CaTa10}, that this problem can be relaxed and solved
efficiently via convex optimization.
 The relaxation of the initial problem can be written as follows: 
\begin{align}
\min_{\bm{A}} \ &\Vert\bm{A}\Vert_* \label{eq:matcompconv1} \\ 
 s.t. \, \, & A_{ij} = X_{ij}, \ \forall i,j \in \Omega,
\label{eq:matcompconv2}
\end{align}
where $\Vert\bm{A}\Vert_*$ denotes the nuclear norm of the matrix $\bm{A}$
with definition: $\Vert \bm{A}\Vert_*=\sum_{k=1}^{\min(m,n)}\sigma_k(\bm{A})$,
with $\sigma_k(\cdot)$ being the $k$--th larger singular value.
This model can be further generalized so that to take into account the presence of 
noise. In that case the equality constraint can be relaxed and the optimization 
problem becomes:
\begin{align}
\min_{\bm{A}} \ \lambda_1 &\Vert\bm{A}\Vert_* + \frac{1}{2}\Vert P_{\Omega} (\bm{A}-\bm{X})\Vert_F^2,  
\end{align}
where   $P_{\Omega}$ is an operator
which sets the entries of its matrix argument not in $\Omega$ to zero, and keeps
the rest unchanged and  $\lambda_1>0$ is a regularization term.

Low rank implies the linear dependence of rows/columns of $\bm{X}$.
 However, this dependence is unstructured.
In many situations, the rows and/or columns of matrix $\bm{X}$ possess additional structure that can
be incorporated into the completion problem in the form of a regularization. In this paper, we assume
that the  rows of $\bm{X}$ are given on vertices of graphs.  
More formally, let us be given an  undirected graph 
$\mathcal{G} = (\bm{V},\bm{E}, \bm{W})$ on the rows with vertices
$\bm{V}= \{1, \ldots , m\}$, edges $\bm{E} \subseteq \bm{V}\times\bm{V}$
and non-negative weights on the edges represented by
the symmetric $m\times m$ matrix $\bm{W}$. If there is an edge between $i,j$, then $W_{ij}=W_{ji}=0$,
and we shall assume the graph has no parallel edges or loops. The latter means that the diagonal
elements of $\bm{W}$ are zero.

The weights capture a strength of association 
between the row elements. 
We embed the graph structure into the matrix completion problem using the \emph{Laplacian}. This
is the positive semidefinite (PSD) matrix $\bm{L}$ defined as $\bm{D}-\bm{W}$ where $\bm{D}$ is the diagonal matrix such that 
$D_{ii}=\sum_{j=1}^m W_{ij}$.

The problem of matrix completion over graphs can be  formulated as follows, \cite{KaBrBrVa14}:
\begin{align}
\min_{\bm{A}} \ \lambda_1 &\Vert\bm{A}\Vert_*  + \frac{1}{2}\Vert P_{\Omega} (\bm{A}-\bm{X})\Vert_F^2
+ \lambda_2\tr\left(\bm{A}^T\bm{L}\bm{A}\right),  \label{eq:matcompconv3} 
\end{align}
where $\tr\left(\bm{A}^T\bm{L}\bm{A}\right)$ is a graph smoothing regularization constraint
 and $\lambda_2 > 0$ is the regularization parameter associated with it. 
In fact it holds that
\begin{equation*}
\sum_{i,j} W_{ij}\Vert \bm{a}_i - \bm{a}_{j}\Vert^2 =\mathrm{tr}\left( \bm{A}^T\bm{L}\bm{A}  \right),
\end{equation*}
with $\bm{a}_i$ being the $i$--th row of the matrix $\bm{A}$.
In words we demand that the rows corresponding
to  neighboring nodes to be ``close" (in some sense) to each other.
This problem, which was originally  been proposed in \cite{KaBrBrVa14} has been generalized
in \cite{shahid2015fast} to tackle scenarios where outliers are present. 

Before we turn our focus to the online problem, we present some useful properties
of the nuclear norm. 
The nuclear norm of a matrix $\bm{M}$ of rank $r$ can be written as \cite{recht2010guaranteed}
 \begin{equation}
\Vert \bm{M} \Vert_*  = \min_{\bm{U}\in\mathbb{R}^{m\times r},\bm{R}\in \mathbb{R}^{r\times n}} \{ \Vert\bm{U}\Vert_F^2 + \Vert\bm{R}\Vert_F^2\}\quad s.t. \, \bm{M} =\bm{U}\bm{R}.
\label{eq:matcompapr} 
\end{equation}
Note that the number of columns of the matrix $\bm{U}$, denoted by $r$, is also a variable.
The problem of estimating $r$ goes beyond the scope of this paper \emph{\textbf{and from now on we will consider that
$r$ will be equal to the rank of $\bm{X}$ and will be known}}. This assumption was also made in other papers
(e.g., \cite{rpca,giannakis}) dealing with online matrix completion. 
Taking this into account and substituting \eqref{eq:matcompapr} into \eqref{eq:matcompconv3} leads us to:
\begin{align}
\min_{\bm{U}, \bm{R}\, : \, \bm{UR} \in \mathbb{R}^{m \times n}} \,&   \lambda_1\left(\Vert\bm{U}\Vert_F^2 + \Vert\bm{R}\Vert_F^2\right)   + \frac{1}{2}\Vert P_{\Omega} (\bm{UR}-\bm{X})\Vert_F^2 \nonumber\\
&+ \lambda_2\tr\left(\bm{R}^T\bm{U}^T\bm{L}\bm{UR}\right).\label{eq:dec1}
\end{align}

\subsection{ONLINE MATRIX COMPLETION ON GRAPHS}
The above deals with the batch problem, i.e., the 
one in which all the measurements are available a priori and are 
used in the computations as a whole.
However, in many applications, having access to all the data may be impractical and/or
infeasible. 
More specifically,   in big data applications, the
data might not be able to be stored  and the  algorithm
needs to retrieve them from slow memory
devices or to access them over networks.
Moreover, in batch operation the unknown subspace has to be re-computed from scratch
whenever a new datum becomes available. 
Our goal here is to present an online solution to the matrix completion over graphs problem. 

In our context we consider that at each step, i.e., $t$, a single
 column of the matrix $\bm{X}$, say $\bm{x}_t\in\Real^m$, which also has missing entries, 
 becomes available. 

Per \eqref{eq:dec1}, each observation
vector $\bm{x}_t\in\Real^m, \ t=1,\ldots, n $ is given by
\begin{equation}
\bm{x}_t = P_{\Omega_t}\left(\bm{U}\bm{r}_t  + \bm{v}_t\right),
\end{equation}
where  $\bm{U}$ is an $m\times r$ matrix $\bm{r}_t\in\Real^r$ and $\bm{v}_t\in \Real^m$ 
is the noise process. This formula will be our starting point for the derivation of the 
online algorithm. Following the exponentially weighted least squares rationale, 
 the online formulation of   \eqref{eq:dec1}  can be cast as follows:
\begin{align}
\min_{\bm{U},\{\bm{r}_{\tau}\}}\sum_{\tau=1}^t & \Big( \frac{1}{2} \Vert P_{\Omega_{\tau}}(\bm{x}_\tau - \bm{U}\bm{r}_{\tau})   \Vert_2^2+\frac{\lambda_1}{2}\Vert\bm{r}_\tau\Vert_2^2  \nonumber\\ 
&+\frac{\lambda_2}{2}(\bm{r}_\tau^T\bm{U}^T\bm{L}\bm{U}\bm{r}_{\tau})\Big)+ \frac{\lambda_1}{2} \Vert\bm{U}\Vert_F^2,\label{eq:prob1}
\end{align}
We attempt to solve the above iteratively. In each iteration $t$ , we maintain the last estimate $\bm{U}_{t-1}$ of the subspace. We compute an optimal $\bm{r}_t$ assuming $\bm{U}_{t-1}$. We then use $\bm{r}_1, \ldots, \bm{r}_t$ to generate $\bm{U}_t$. This two-step procedure is
typical in online matrix factorization problems, see for example \cite{Mairal,slavakis2014modeling}. 
Next we derive the minimization for the first step of the algorithm. To that end, we keep only
the terms which depend on $\bm{r}$ and we obtain: 
\begin{equation}
\min_{\bm{r}} \frac{1}{2} \Vert P_{\Omega_{t}}(\bm{x}_t - \bm{U}_{t-1}\bm{r})   \Vert_2^2  +  \frac{\lambda_1}{2}\Vert\bm{r}\Vert_2^2+\frac{\lambda_2}{2}(\bm{r}^T\bm{U}^T_{t-1}\bm{L}\bm{U}_{t-1}\bm{r}).\label{eq:probr}
 \end{equation}
Computing the derivative with respect to $\bm{r}$ and setting it equal to  $\bm{0}_r$ we obtain
the minimizer of \eqref{eq:probr} given by:
\begin{equation}
\bm{r}_t = \bm{A}_t^{-1} \bm{U}_{t-1}^TP_{\Omega_t}(\bm{x}_t),
\label{eq:updr}
\end{equation}
where 
\[
\bm{A}_t = \lambda_1\bm{I}_r + \bm{U}_{t-1}^T(\bm{\Omega}_t+\lambda_2\bm{L})\bm{U}_{t-1}
\]
and $\bm{\Omega}_t\in\Real^{m\times m}$ is the diagonal matrix associated with the set $\Omega_t$ having in its diagonal $1$ if the respective entry is observed and $0$ if it is unobserved.
Note that $\lambda_1$ being positive implies $\bm{A}_t$ is positive definite and therefore invertible. 

The next step  is  the minimization with respect to $\bm{U}$.
Computing the gradient of \eqref{eq:prob1} with respect to $\bm{U}$, and equating it with the zero matrix, we obtain:
\begin{align}
\lambda_1 \bm{U} +\lambda_2 \bm{L}\bm{U}\bm{R}_t+\sum_{\tau=1}^t \bm{\Omega}_{\tau} \bm{U} \bm{r}_{\tau}\bm{r}_{\tau}^T =\bm{P}_t \label{eq:optmis} 
\end{align}
where 
\begin{align}
\bm{R}_t  &= \sum_{\tau=1} ^t \bm{r}_{\tau}\bm{r}_{\tau}^T \label{eq:updRt}\\
\bm{P}_t &=  \sum_{\tau=1}^t\bm{\Omega}_\tau \bm{x}_\tau\bm{r}_\tau^T \label{eq:updPt}.
\end{align}
A drawback of this formulation is that the matrix $\bm{U}$ is 
coupled with $\bm{\Omega}_{\tau}$ and $\bm{r}_{\tau}$ so solving directly \eqref{eq:optmis}
with respect to $\bm{U}$ becomes difficult or infeasible. However, we can use 
properties of Kronecker products 
and bypass this difficulty.
First, we vectorize \eqref{eq:optmis} and we obtain:
\begin{align*}
&\vect\left\lbrace\sum_{\tau=1}^t \bm{\Omega}_{\tau} \bm{U} \bm{r}_{\tau}\bm{r}_{\tau}^T\right\rbrace +\lambda_1 \vect\left\lbrace\bm{U}\right\rbrace +\lambda_2 \vect\left\lbrace\bm{L}\bm{U}\bm{R}_t\right\rbrace\\
& \qquad=\vect\left\lbrace\bm{P}_t\right\rbrace,
\end{align*}
where $\vect\{\}$ is the vectorization operator that vectorizes a matrix by stacking the 
columns so as to form a supervector.
The first term of the left hand side can be equivalently written \cite{laub2005matrix}:
\begin{equation}
\vect\left\lbrace\sum_{\tau=1}^t \bm{\Omega}_{\tau} \bm{U} \bm{r}_{\tau}\bm{r}_{\tau}^T\right\rbrace 
= \left( \sum_{\tau=1}^t \bm{r}_{\tau}\bm{r}_{\tau}^T \otimes \bm{\Omega}_{\tau}\right) \bm{u},
\end{equation}
where $\bm{u}:=\vect\lbrace\bm{U}\rbrace$. 
The third term of the left hand side of \eqref{eq:optmis} can be  written  as:
\begin{equation}
\vect\left\lbrace\bm{L}\bm{U}\bm{R}_t\right\rbrace = \left(\bm{R}_t\otimes\bm{L} \right)\bm{u}
\end{equation}
So, the solution of \eqref{eq:optmis} (in a vectorized form) is given by:
\begin{equation}
\bm{u} = \left( \sum_{\tau=1}^t \bm{r}_{\tau}\bm{r}_{\tau}^T \otimes \bm{\Omega}_{\tau} +\lambda_1\bm{I}_{mr} + \bm{R}_t\otimes\bm{L} \right)^{-1}\bm{p}_t,
\label{eq:solmisentr}
\end{equation}
where $\bm{p}_t = \vect\lbrace\bm{P}_t\rbrace$.
The steps of the algorithm are summarized as Algorithm \ref{alg:SpR}

\begin{algorithm}
	\DontPrintSemicolon
		\KwIn{$\lambda_1$, $\lambda_2$, $\bm{L}$}
	\KwOut{Computed Subspaces $ {\bm{U}}_t$ and vectors $\bm{r}_t$}
\textbf{Initialize: } $\bm{U}_0$\; 
\For{ $t=1,2,\ldots$}
{Compute $\bm{r}_t$ by solving \eqref{eq:updr}\;
Update $\bm{R}_t, \ \bm{P}_t$ by \eqref{eq:updRt} \eqref{eq:updPt} respectively \label{UpdateLine}\;
Compute $ {\bm{U}}_t$ by solving  \eqref{eq:solmisentr} and devectorizing
 }
 \caption{Online Matrix Completion on Graphs}
	\label{alg:SpR}
\end{algorithm}

A crucial point regarding the computational aspects of Algorithm \ref{alg:SpR} is that the memory and time complexities do not grow with
time: The update in line \ref{UpdateLine} only needs
to use the previous values $\bm{R}_{t-1,} \ \bm{P}_{t-1}$, and the new quantities $\bm{r}_t,\bm{x}_t, \bm{\Omega}_t$. Hence, only
a bounded amount of memory (and computation time) is required. 

\subsubsection*{Full Observability}
In the special case where the entries of each $\bm{x}_t$ are fully observable (and so $\bm{\Omega}_t$ becomes the identity matrix), we can take a more direct approach. This is the \emph{subspace tracking problem}.
Since, $\lambda_2>0$ and $\bm{L}$ is positive semidefinite, from \eqref{eq:optmis} we can write 
\begin{align*}
 \lambda_1(\bm{I}_m+\lambda_2 \bm{L})^{-1} \bm{U}  + \bm{U}\bm{R}_t= (\bm{I}_m+\lambda_2 \bm{L})^{-1}\bm{P}_t. 
\end{align*}

This belongs to the family of the so-called \emph{Sylvester's equations} (see, e.g., \cite{sylvester}),
and can be solved efficiently. 
The general form of Sylvester's equation is:
\begin{equation*}
\bm{A}\bm{X} + \bm{X}\bm{B} =\bm{C},
\end{equation*}
and has a unique solution when there are no common eigenvalues of $\bm{A}$ and $-\bm{B}$. For our case, this is assured because $\bm{R}_t$ is PSD. 

%

 \section{ROBUSTIFICATION}\label{ROSTG}
 A drawback of the matrix completion techniques,
 which rely on the Frobenious  norm minimization
 is that they are sensitive
 to heavy tailed noise. In the batch scenario, 
 Robust PCA (RPCA) originally proposed in 
 \cite{candes2011robust} overcomes this limitation. In particular, 
 the  model generating
 the  matrix comprising missing entries is the following:
 \begin{equation}
\bm{M} = \bm{A} + \bm{S},
\end{equation}  
 where $\bm{A}$ is a low rank matrix and $\bm{S}$ is a sparse matrix,
 the entries of which have arbitrarily large amplitude;
 the latter matrix models the outlier noise.
 The optimization problem for the
  matrix completion takes the following form:
  \begin{align*}
  \min_{\bm{A},\bm{S}}  &\Vert\bm{A}\Vert_* +\lambda_s \Vert\bm{S}\Vert_1,\\   
 &\mathrm{s.t.} \,\,  \bm{M} = \bm{A}+\bm{S},
  \end{align*}
   where $\Vert\cdot\Vert_1$ promotes sparsity and has the following definition
   $\Vert\bm{S}\Vert_1 = \sum_{i,j} \left|S_{i,j}\right|$, i.e., the sum of absolute values
	of the entries of $S$.

 The aforementioned problem has been also extended to the online scenario, e.g., \cite{rpca,giannakis}.
 This will be our starting point for deriving the online robust MC algorithm on graphs.
 To be more specific, the model generating the columns of the matrix becomes:
 \begin{equation*}
 \bm{x}_t = P_{\Omega_t} \left( \bm{U}\bm{r}_t  + \bm{s}_t +\bm{v}_t\right),
 \end{equation*}
 where $\bm{s}_t$ stands for the outlier vector. For this, we assume that the $\ell_0$ (pseudo) norm,
which counts the number of non-zero coefficients,
 is bounded and smaller than $m$, i.e., $\Vert\bm{s}_t \Vert_0 \leq m' < m$.\footnote{In practice if $m'=O(\log m)$ then we can recover the sparse vector.}
 Furthermore, similarly to what we have done before, we assume that there exists an
 underlying graph structure, which is assumed to be known.
 The problem we want to solve becomes:
 \begin{align}
&\min_{\bm{U},\{\bm{r}_{\tau}\},\{\bm{s}_{\tau}\}}\frac{\lambda_1}{2} \Vert\bm{U}\Vert_F^2+\sum_{\tau=1}^t \Big( \frac{1}{2}\Vert P_{\Omega_{\tau}}(\bm{x}_\tau - \bm{U}\bm{r}_{\tau}- \bm{s}_\tau) \Vert_2^2 \nonumber\\
&\qquad \qquad \qquad+\frac{\lambda_1}{2}\Vert\bm{r}_\tau\Vert_2^2+\frac{\lambda_2}{2}(\bm{r}_\tau^T\bm{U}^T\bm{L}\bm{U}\bm{r}_{\tau})+\frac{\lambda_3}{2} \Vert \bm{s}_\tau \Vert_1 \Big),\label{eq:mainObjective}
 \end{align}
where $\lambda_3>0$.

 For given $\bm{\Omega}, \bm{U}$, $\bm{x}$ and $\bm{s}$, the expression 
\begin{equation}
\Vert \bm{\Omega}(\bm{x} - \bm{U}\bm{r}- \bm{s}) \Vert_2^2 +\lambda_1\Vert\bm{r}\Vert_2^2+\lambda_2(\bm{r}^T\bm{U}^T\bm{L}\bm{U}\bm{r})+\lambda_3 \Vert \bm{s} \Vert_1 \label{eq:rsMin}
\end{equation}
is minimized when 
\begin{equation}
\bm{{r}}=\bm{B}(\bm{x}-\bm{s}), \label{eq:updrFull}
\end{equation}
where 
\[
\bm{B}=\bm{A}^{-1}\bm{U}^T\bm{\Omega}
\]
 and 
\[
\bm{A}=\lambda_1\bm{I}_r+\bm{U}^T(\bm{\Omega}+\lambda_2\bm{L})\bm{U}.
\]
Treating $\bm{r}$ as a function of $\bm{s}$ and plugging it back into \eqref{eq:rsMin}, 
the joint minimization of $\bm{r}$ and $\bm{s}$ for given $\bm{U}$ is formulated as
\begin{align*}
\min_{\bm{s}}\, &\Vert \bm{\Omega}(\bm{I}_m - \bm{U}\bm{B})(\bm{x}-\bm{s}) \Vert_2^2 +\Vert\sqrt{\lambda_1}\bm{B}(\bm{x}-\bm{s})\Vert_2^2\\
&+\Vert\sqrt{\lambda_2} \bm{L}^{\frac{1}{2}}\bm{U}\bm{B}(\bm{x}-\bm{s})\Vert_2^2+\lambda_3 \Vert \bm{s} \Vert_1,
\end{align*}
where we have used that $\bm{L}$ is PSD. This, in turn can be formulated as the following lasso estimator: 
\begin{equation}
\min_{\bm{s}}\, \Vert \bm{C}(\bm{x}-\bm{s}) \Vert_2^2 +\lambda_3 \Vert \bm{s} \Vert_1, \label{lasso}
\end{equation}
where $\bm{C}$ is the $(m+r+m)\times m$ matrix such that 
\[
\bm{C}=\left[\left(\bm{\Omega}(\bm{I}_m - \bm{U}\bm{B})\right)^T,\sqrt{\lambda_1}\bm{B}^T, \sqrt{\lambda_2} \left(\bm{L}^{\frac{1}{2}}\bm{U}\bm{B}\right)^T\right]^T.
\]

This is a convex optimization problem and therefore efficiently solvable. We use the above to compute $\bm{r}_t$ and $\bm{s}_t$ using
$\bm{\Omega}_t$, $\bm{x}_t$ and $\bm{U}_{t-1}$. Similar to the above algorithm, we use these computed values to compute $\bm{U}_t$.

Taking partial derivative of \eqref{eq:mainObjective} with respect to $\bm{U}$ and setting it to zero, we get 
\[
\bm{Q}_t=\lambda_1\bm{U}+\lambda_2\bm{LUR}_t+\sum_{\tau=1}^t\bm{\Omega}_\tau\bm{U} \bm{r}_\tau\bm{r}_\tau^T
\]
where $\bm{Q}_t=\sum_{\tau=1}^t\bm{\Omega}_\tau(\bm{x}_\tau-\bm{s}_\tau)\bm{r}_\tau^T$ and, as before, $\bm{R}_t = \sum_{\tau=1}^t \bm{r}_\tau\bm{r}_\tau^T$. 

As before, we vectorize and solve, thereby getting
\begin{equation}
\bm{u} = \left( \sum_{\tau=1}^t \bm{r}_{\tau}\bm{r}_{\tau}^T \otimes \bm{\Omega_{\tau}} +\lambda_1\bm{I}_{mr} + \bm{R}_t\otimes\bm{L} \right)^{-1}\bm{q}_t, \label{eq:UUpdate3}
\end{equation}
where $\bm{q}_t=\vect\{\bm{Q}_t\}$.

The algorithm is summarized as Algorithm \ref{robustAlg}.

\begin{algorithm}
	\DontPrintSemicolon
		\KwIn{$\lambda_1$, $\lambda_2$, $\bm{L}$ }
\KwOut{Computed Subspaces $ {\bm{U}}_t$ and vectors $\bm{r}_t, \ \bm{s}_t$}
\textbf{Initialize: } $\bm{U}_0$\; 
\For{ $t=1,2,\ldots$}
{Compute $\bm{s}_t$ by solving the lasso \eqref{lasso} using $\bm{U}_{t-1}$ and $\bm{\Omega}_t$\;
Compute $\bm{r}_t$ by applying equation \eqref{eq:updrFull}\;
Update $\bm{R}_t$ and $\bm{Q}_t$ using $\bm{x}_t$, $\bm{r}_t$ and $\bm{s}_t$\;
Compute $\bm{U}_t$ using\eqref{eq:UUpdate3}
 \caption{Online Robust Matrix Completion on Graphs}
 }
	\label{robustAlg}
\end{algorithm}

\section{CONVERGENCE}
In this section we will discuss the convergence of the proposed scheme, in particular, the robust scheme with missing entries.
The convergence proofs for the other schemes follow similar steps.
 
Define the following:
$g_{t}(\bm{U}, \bm{r}, \bm{s}) :=   \Big( \frac{1}{2}\Vert P_{\Omega_{t}}(\bm{x}_t - \bm{U}\bm{r}- \bm{s})\Vert_2^2 + \frac{\lambda_1}{2}\Vert\bm{r} \Vert_2^2 +\frac{\lambda_2}{2}(\bm{r}^T\bm{U}^T\bm{L}\bm{U}\bm{r}) + \lambda_3 \Vert \bm{s} \Vert_1 \Big),$ and
$g_{t}(\bm{U}) := \min_{\bm{r}, \bm{s}}  g_{t}(\bm{U}, \bm{r}, \bm{s}).$ The proposed algorithm effectively aims to minimize the following\footnote{We normalize with $t$ so as to prevent the existence of unbounded values. It can be readily seen that the solution
at each time step doesn't depend on the normalization.}:
\begin{equation*}
C_t(\bm{U}) = \frac{1}{t} \sum_{\tau=1}^t g_{\tau}(\bm{U}) + \frac{\lambda_1}{2t} \Vert\bm{U}\Vert_F^2.
\end{equation*}

It is worth pointing out that as time increases, minimization of $C_t(\bm{U})$ becomes computationally demanding
since it involves solving $t$ least squares and $t$ $\ell_1$ minimization problems for the estimation
of $\bm{r}$ and $\bm{s}$ respectively. For this reason, the algorithm actually minimizes the following approximation of the above
cost function:
\begin{equation}
\Capr(\bm{U}) = \frac{1}{t} \sum_{\tau=1}^t g_{\tau}(\bm{U},\bm{r}_{\tau},\bm{s}_{\tau}) + \frac{\lambda_1}{2t} \Vert\bm{U}\Vert_F^2,
\end{equation}
where 
\[
\{\bm{r}_{t},\bm{s}_{t}\}=\arg\min_{\bm{r},\bm{s}}g_{t}(\bm{U}_{t-1},\bm{r},\bm{s}).
\]
For the analysis of convergence, we make the following assumptions:
\begin{itemize}
\item \textbf{A1:} $\{\Omega_t\}_t$ and $\{\bm{x}_t\}_t$ are i.i.d. random processes.
\item \textbf{A2:} Each $\bm{x}_t$ and $\bm{U}_t$ is in fixed compact set $\mathcal{X} \subset \Real^{m}$ and
$\mathcal{C}\subset\Real^{m\times r}$, respectively.
\item \textbf{A3:} $\Capr(\bm{U})$ is strongly convex, i.e., 
$\lambda_{\min}(\nabla^2\Capr(\bm{U}))\geq \epsilon$ for a positive constant $\epsilon$.
\item \textbf{A4:} The lasso given in equation \eqref{lasso} has a unique solution. 
\end{itemize}
Before we proceed to the proof a few words on the assumptions are due.
Assumption A1 is typically adopted in several online learning problems,
e.g., \cite{Th15},
and has been made in the online matrix completion problem, e.g., \cite{giannakis}.
For $\bm{x}_t$, A2 naturally holds in many applications, e.g., media, data transmission. For $\bm{U}_t$ is a technical assumption which simplifies the 
proof and has been verified through extensive simulations; however, it is also reasonable in many cases to suppose that the principle vectors of an underlying subspace are bounded. This is
especially the case where the application forces it, e.g., you can only rate a movie one to five stars.
Regarding assumption A3, we assume that the Hessian of the cost function is bounded. This is also considered
in \cite{Mairal,giannakis} and essentially implies that $\frac{1}{t}\left( \sum_{\tau=1}^t \bm{r}_{\tau}\bm{r}_{\tau}^T \otimes \bm{\Omega}_{\tau} +\lambda_1\bm{I}_{mr} + \bm{R}_t\otimes\bm{L} \right)\succcurlyeq \epsilon\bm{I}_{mr}$. 
An additional regularization term can be added to ensure that this assumption holds, but here for simplicity we won't consider such a case.
Assumption A4 is reasonable since it is helped by  the uniformly random matrices $\bm{\Omega}_t$ not affecting too much the incoherence of the subspace estimates $\bm{U}_{t-1}$.   

We wish to show the following:
\begin{thm}\label{thm:convergence}
If assumptions A1 -- A4 hold, then Algorithm \ref{robustAlg} converges to a stationary point of the objective
function, i.e., $\lim_{t\to\infty} \nabla C_t(\bm{U}_t) = \bm{O}_{r\times m}$.
\end{thm}
In a nutshell, this theorem states that asymptotically the estimated subspace
minimize the \textit{original} cost function, despite the fact that the 
estimates occur
from the minimization of an \textit{approximate} cost function. 

Mardani et al.~\cite{giannakis} study an online matrix completion-type problem in the context of tracking network anomalies. Application aside, 
and framed in our notation, the algorithms they present essentially try to compute the same low-rank $\bm{U}_t$, matrices and sparse vectors $\bm{s}_t$
as our algorithms do, but they make no use of graph structure in the sense we have done via the Laplacian. They prove a version of Theorem \ref{thm:convergence},
but for us to apply their proof technique (which is, in turn, based on \cite{Mairal}), we must ensure the following lemma holds:
\begin{lem}\label{lem:Lipschitz}
If assumptions A2 and A4 hold, then for $\bm{U}$ in a compact set $\mathcal{C}\subset\Real^{m\times r}$, the following are Lipschitz continuous functions of $\bm{U}$ with constants independent
of $t$: (i) $\{\bm{r}_{t}(\bm{U}),\bm{s}_{t}(\bm{U})\}=\arg\min_{\bm{r},\bm{s}}g_{t}(\bm{U},\bm{r},\bm{s})$, (ii) $g_t(\bm{U}, \bm{r}, \bm{s})$, for fixed $\bm{r}, \bm{s}$, (iii) $g_t(\bm{U})$, 
and (iv) $\nabla g_t(\bm{U})$.
\end{lem} 
Lemma \ref{lem:Lipschitz} above is the equivalent of Lemma 1 in \cite{giannakis}, and we modify their proof to cope with the terms arising from the Laplacian.
 Define
\begin{align*}
\bm{M}_t(\bm{U}) := \Big[&\bm{\Omega}_t(\bm{I}_m - \bm{U}\bm{B}_t(\bm{U}))\\
&\sqrt{\lambda_1}\bm{B}_t(\bm{U})\\
&\sqrt{\lambda_2} \bm{L}^{\frac{1}{2}}\bm{U}\bm{B}_t(\bm{U})\Big],
\end{align*}
where $\bm{B}_t(\bm{U}) := \bm{A}_t(\bm{U})^{-1}\bm{U}^T\bm{\Omega}_t$ and $\bm{A}_t(\bm{U}) :=\lambda_1\bm{I}_r+\bm{U}^T(\bm{\Omega}_t+\lambda_2\bm{L})\bm{U}$. Note,
$\bm{A}_t(\bm{U})$ is positive definite, and therefore invertible. 

For simplicity, we omit the subscript $t$ below where it does not aid the argument.  

\emph{Proof of Lemma \ref{lem:Lipschitz}}
(i) 
As in Section \ref{ROSTG}, $\bm{r}$ can first be expressed as an affine function of $\bm{s}$ (see \eqref{eq:updrFull}), and after the Lipschitz continuity of 
$\bm{s}(\bm{U})$ is demonstrated, the Lipschitz continuity of $\bm{r}(\bm{U})$ follows easily. This is the approach taken in \cite{giannakis}, and we apply a modified
version of it below. Thus, defining 
\[
u(\bm{s},\bm{U}_1, \bm{U}_2) := \Vert \bm{M}(\bm{U}_1)\left(\bm{x}-\bm{s}\right) \Vert_2^2 - \Vert \bm{M}(\bm{U}_2)\left(\bm{x}-\bm{s}\right) \Vert_2^2
\]
(cf. \eqref{lasso}), it is shown that 
\[
u(\bm{s}(\bm{U}_2),\bm{U}_1, \bm{U}_2)-u(\bm{s}(\bm{U}_1),\bm{U}_1, \bm{U}_2) \geq c_0 \Vert \bm{s}(\bm{U}_2) -\bm{s}(\bm{U}_1)  \Vert_2^2 
\]
for some constant $c_0>0$ independent of $t$. This holds for our case as well. It is then shown that $u(., \bm{U}_1, \bm{U}_2)$ is Lipschitz continuous, and we can follow the same steps of the proof
until it is required to show that $\bm{M}(\bm{U})$ is Lipschitz continuous. Here we have to cater for the terms related to the Laplacian. It is quite possible to apply the same technique as in 
\cite{giannakis},  but we use a shorter argument thus:
\begin{align}
&\Vert \bm{M}(\bm{U}_1) - \bm{M}(\bm{U}_2) \Vert \leq \Vert \bm{\Omega}[\bm{U}_1\bm{B}(\bm{U}_1) - \bm{U}_2\bm{B}(\bm{U}_2)] \Vert \nonumber\\
&+\sqrt{\lambda_1} \, \Vert \bm{B}(\bm{U}_1) - \bm{B}(\bm{U}_2) \Vert + \sqrt{\lambda_2} \, \Vert \bm{L}^{\frac{1}{2}}[\bm{U}_1\bm{B}(\bm{U}_1)-\bm{U}_2\bm{B}(\bm{U}_2)]\Vert \nonumber \\
& \leq \left(1+\sqrt{\lambda_2} \Vert \bm{L}^{\frac{1}{2}} \Vert\right) \Vert \bm{U}_1\bm{A}(\bm{U}_1)^{-1}\bm{U}_1^T - \bm{U}_2\bm{A}(\bm{U}_2)^{-1}\bm{U}_2^T\Vert \nonumber \\
& +  \sqrt{\lambda_1} \, \Vert \bm{A}(\bm{U}_1)^{-1}\bm{U}_1^T - \bm{A}(\bm{U}_2)^{-1}\bm{U}_2^T\Vert.\label{Lipbound}
\end{align}
Now consider the function $f(\bm{U}) := \bm{A}(\bm{U})^{-1}\bm{U}^T$. This is differentiable with respect to $\bm{U}$ and since $\bm{U}$ is assumed to be constrained to a fixed compact
space $\mathcal{C}\subset\Real^{m\times r}$, Lipschitz continuity of $f(\bm{U})$ follows by the mean value theorem. Similarly for $\bm{U}f(\bm{U})$. It follows that there is a constant $c_2>0$
 independent of $t$ such  \eqref{Lipbound} is bounded by $c_2\Vert\bm{U}_1-\bm{U}_2\Vert$. The rest of the rest of the proof for part (i) follows as in \cite{giannakis}. 

(ii) $g_t(\bm{U},\bm{r}, \bm{s})$ is a quadratic function of $\bm{U}$ on a compact set and so clearly Lipschitz.

(iii) Using $g_t(\bm{U})=g_t(\bm{U}, \bm{r}_{t}(\bm{U}),\bm{s}_{t}(\bm{U}))$ where $\{\bm{r}_{t}(\bm{U}),\bm{s}_{t}(\bm{U})\}=\arg\min_{\bm{r},\bm{s}}g_{t}(\bm{U},\bm{r},\bm{s})$,
we have (omitting $t$ subscripts) 
\begin{align*}
&g(\bm{U}_2)-g(\bm{U}_1) =  \frac{1}{2}\Vert P_{\Omega}(\bm{U}_2\bm{r}(\bm{U}_2)+ \bm{s}(\bm{U}_2))\Vert_2^2\\
&-\frac{1}{2}\Vert P_{\Omega}(\bm{U}_1\bm{r}(\bm{U}_1)+ \bm{s}(\bm{U}_1))\Vert_2^2\\
&+\langle \Omega\bm{x},\,   \bm{U}_1\bm{r}(\bm{U}_1)+ \bm{s}(\bm{U}_1) - \bm{U}_2\bm{r}(\bm{U}_2) -\bm{s}(\bm{U}_2)\rangle \\
&+\frac{\lambda_1}{2}\left(\Vert \bm{r}(\bm{U}_2)\Vert_2^2 -\Vert \bm{r}(\bm{U}_1)\Vert_2^2 \right)+\lambda_3\left(\Vert \bm{s}(\bm{U}_2)\Vert_1 -\Vert \bm{s}(\bm{U}_1)\Vert_1 \right)\\
&+\frac{\lambda_2}{2}\left(\bm{r}(\bm{U}_2)^T\bm{U}_2^T\bm{L}\bm{U}_2\bm{r}(\bm{U}_2) - \bm{r}(\bm{U}_1)^T\bm{U}_1^T\bm{L}\bm{U}_1\bm{r}(\bm{U}_1)  \right).
\end{align*}
As demonstrated in \cite{giannakis}, the first term is bounded as 
\begin{align*}
&\Vert P_{\Omega}(\bm{U}_2\bm{r}(\bm{U}_2)+ \bm{s}(\bm{U}_2))\Vert_2^2-\Vert P_{\Omega}(\bm{U}_1\bm{r}(\bm{U}_1)+ \bm{s}(\bm{U}_1))\Vert_2^2\\
& \leq c_3 \Big(\Vert \bm{U}_2 - \bm{U}_1\Vert \, \Vert \bm{r}(\bm{U}_2) \Vert_2 + \Vert \bm{U}_1 \Vert \, \Vert\bm{r}(\bm{U}_2)-\bm{r}(\bm{U}_1) \Vert_2\\
&\qquad \quad+\Vert\bm{s}(\bm{U}_2)-\bm{s}(\bm{U}_1) \Vert_2 \Big)
\end{align*}
for some constant $c_3>0$, the second is bounded as 
\begin{align*}
&\langle \Omega\bm{x},\,   \bm{U}_1\bm{r}(\bm{U}_1)+ \bm{s}(\bm{U}_1) - \bm{U}_2\bm{r}(\bm{U}_2) -\bm{s}(\bm{U}_2)\rangle  \\
&\leq  \Big(\Vert \bm{U}_2 - \bm{U}_1\Vert \, \Vert \bm{r}(\bm{U}_2)\Vert_2 + \Vert \bm{U}_1\Vert \, \Vert \bm{r}(\bm{U}_2)-\bm{r}(\bm{U}_1) \Vert_2 \\
&\qquad +\Vert \bm{s}(\bm{U}_2)-\bm{s}(\bm{U}_1) \Vert_2  \Big)\Vert P_{\Omega}(\bm{x})\Vert_2, 
\end{align*}
and the third term is bounded as 
\begin{align*}
&\frac{\lambda_1}{2}\left(\Vert \bm{r}(\bm{U}_2)\Vert_2^2 -\Vert \bm{r}(\bm{U}_1)\Vert_2^2 \right)+\lambda_3\left(\Vert \bm{s}(\bm{U}_2)\Vert_1 -\Vert \bm{s}(\bm{U}_1)\Vert_1 \right)\\
&\leq \frac{\lambda_1}{2}\Vert \bm{r}(\bm{U}_2)- \bm{r}(\bm{U}_1)\Vert_2\left(\Vert \bm{r}(\bm{U}_2)\Vert_2 +\Vert \bm{r}(\bm{U}_1)\Vert_2\right)\\
&\quad + \lambda_3\sqrt{r}\Vert \bm{s}(\bm{U}_2)- \bm{s}(\bm{U}_1)\Vert_2.
\end{align*}
By previous results, all the above terms are Lipschitz continuous. This was shown in \cite{giannakis}. It remains to show Lipschitz continuity for the final term. 
\begin{align*}
&\bm{r}(\bm{U}_2)^T\bm{U}_2^T\bm{L}\bm{U}_2\bm{r}(\bm{U}_2) - \bm{r}(\bm{U}_1)^T\bm{U}_1^T\bm{L}\bm{U}_1\bm{r}(\bm{U}_1)\\
&=\Vert \sqrt{\bm{L}}\, \bm{U}_2\bm{r}(\bm{U}_2)\Vert_2^2 - \Vert \sqrt{\bm{L}}\, \bm{U}_1\bm{r}(\bm{U}_1)\Vert_2^2\\
&= \left(\Vert \sqrt{\bm{L}}\, \bm{U}_2\bm{r}(\bm{U}_2)\Vert_2 - \Vert \sqrt{\bm{L}}\, \bm{U}_1\bm{r}(\bm{U}_1)\Vert_2\right)\\
&\quad \times \left(\Vert \sqrt{\bm{L}}\, \bm{U}_2\bm{r}(\bm{U}_2)\Vert_2 + \Vert \sqrt{\bm{L}}\, \bm{U}_1\bm{r}(\bm{U}_1)\Vert_2\right).
\end{align*}
By virtue of compactness the last term is bounded from above by some positive constant independent of $t$, and
\begin{align*}
& \Vert \sqrt{\bm{L}}\, \bm{U}_2\bm{r}(\bm{U}_2)\Vert_2 - \Vert \sqrt{\bm{L}}\, \bm{U}_1\bm{r}(\bm{U}_1)\Vert_2\\
&\leq \Vert \sqrt{\bm{L}}\, \bm{U}_2\bm{r}(\bm{U}_1)\Vert_2 - \Vert \sqrt{\bm{L}}\, \bm{U}_1\bm{r}(\bm{U}_1)\Vert_2\\
&\quad  + \Vert \sqrt{\bm{L}}\, \bm{U}_2\left(\bm{r}(\bm{U}_2) - \bm{r}(\bm{U}_1)\right) \Vert_2.
\end{align*}
By the submultiplicativity property of the operator norm, we have 
\[
 \Vert \sqrt{\bm{L}}\, \bm{U}_2\left(\bm{r}(\bm{U}_2) - \bm{r}(\bm{U}_1)\right) \Vert_2 \leq \Vert \sqrt{\bm{L}}\, \bm{U}_2 \Vert \, \Vert \bm{r}(\bm{U}_2) - \bm{r}(\bm{U}_1) \Vert.
\]
Furthermore, 
\begin{align*}
&\leq \Vert \sqrt{\bm{L}}\, \bm{U}_2\bm{r}(\bm{U}_1)\Vert_2 - \Vert \sqrt{\bm{L}}\, \bm{U}_1\bm{r}(\bm{U}_1)\Vert_2\\
&\leq \Vert \sqrt{\bm{L}}\, \bm{U}_2\bm{r}(\bm{U}_1)- \sqrt{\bm{L}}\, \bm{U}_1\bm{r}(\bm{U}_1)\Vert_2\\
&\leq \Vert \sqrt{\bm{L}}\Vert \,\Vert \left(\bm{U}_2-\bm{U}_1\right)\Vert \, \Vert\bm{r}(\bm{U}_1)\Vert_2
\end{align*}
 where the last inequality follows by two applications of the submultiplicativity property of the operator norm. By compactness, 
$\Vert\sqrt{\bm{L}}\, \bm{U}_2 \Vert$ and $\Vert \sqrt{\bm{L}}\Vert \,\Vert\bm{r}(\bm{U}_1)\Vert_2$ are both bounded from above by some positive constant independent of $t$.

Putting it all together proves the Lipschitz continuity of $g_t(\bm{U})$.

(iv) Since by assumption $\{\bm{r}_{t}(\bm{U}),\bm{s}_{t}(\bm{U})\}=\arg\min_{\bm{r},\bm{s}}g_{t}(\bm{U},\bm{r},\bm{s})$ is unique as a minimizer of $g_t(\bm{U}, \bm{r}, \bm{s})$ 
for a given $\bm{U}$, a theorem of Danskin (see, e.g., \cite{Ber99}) allows us to say
\begin{align*}
\nabla g_t(\bm{U}) &= \bm{r}_t(\bm{U})\Big(\bm{U}\bm{r}_t(\bm{U})+\bm{s}_t(\bm{U})-\bm{x}_t(\bm{U})\Big)\bm{\Omega}\\
&\quad + \lambda_2 \bm{r}_t(\bm{U}) \bm{r}_t(\bm{U})^T \bm{U}^T\bm{L}. 
\end{align*}
To prove $g_t(\bm{U})$ is Lipschitz continuous, \cite{giannakis} has already shown $\Vert g_t(\bm{U}_2)- g_t(\bm{U}_1) \Vert_F \leq c_4 \Vert \bm{U}_2  - \bm{U}_1\Vert$ for the first term,
and the proof applies just as well for the above. It remains to deal with the second term. 

Writing $\bm{R}_i$ for $\bm{r}_t(\bm{U}_i) \bm{r}_t(\bm{U}_i)^T$, we have 
\begin{align*}
&\Vert \left(\bm{R}_2\bm{U}_2 - \bm{R}_1\bm{U}_1\right)  \bm{L}\Vert_F \leq \Vert \bm{R}_2\bm{U}_2 - \bm{R}_1\bm{U}_1\Vert_F \, \Vert \bm{L}\Vert_F\\
& \leq \Vert \bm{L}\Vert_F \Big(\Vert \bm{R}_2-\bm{R}_1\Vert_F\, \Vert \bm{U}_2 \Vert_F +   \Vert\bm{R}_1\Vert_F\, \Vert \bm{U}_2 - \bm{U}_1\Vert_F  \Big)
\end{align*}
and from here it is straightforward to see that Lipschitz continuity follows. 
\qedsymbol

Having proved Lemma \ref{lem:Lipschitz}, Lemma \ref{CconvLemma} below can be proved exactly as done in \cite{giannakis} (we omit the proof here). The lemma is used 
in the proof of Theorem \ref{thm:convergence}, summarized below.


\begin{lem}[\cite{giannakis}]
If Assumptions A1 -- A4 hold then $\Capr(\bm{U}_t)$ converges and $\Capr(\bm{U}_t) - C(\bm{U}_t)\to 0$ almost surely. \label{CconvLemma}
\end{lem}

\emph{Proof overview of Theorem \ref{thm:convergence}~\cite{giannakis}:} First, since the $\bm{U}_t$ belong to a compact subset, then one can choose a
convergent subsequence for which $\lim_{t_i\to \infty} \bm{U}_{t_i} = \bm{U}_*$.
With a slight abuse of notation $t_i$ will be substituted by $t$. Choose a sequence $\alpha_t>0$ 
for which $\alpha_t \to 0, \ t\to \infty$. It holds that $\Capr(\bm{U}_t+\alpha_t\bm{U}_o) \geq C_t(\bm{U}_t+\alpha_t\bm{U}_o), \ \forall \bm{U}_o$, since the approximate cost always overestimates $C_t$.
Exploiting the mean value theorem and Lemma \ref{CconvLemma}, it can be shown that
\begin{align}
\lim_{t\to\infty} &\tr(\bm{U}_o^T(\nabla\Capr(\bm{U}_t)-\nabla C_t(\bm{U}_t)))  \nonumber \\
&+ \lim_{t\to\infty} \frac{1}{2} \alpha_t \tr(\bm{U}_o^T(\nabla^2 \Capr(\bm{U}^1_t)-\nabla^2 C(\bm{U}^2_t))) \geq 0,
 \end{align}
for some matrices $\bm{U}^1_t, \ \bm{U}^2_t$. It can be readily shown that the second term tends to zero,
since all the involved quantities apart from $\alpha_t$ are bounded and $\alpha_t\to 0$.
So, we have that 
\begin{align}
\lim_{t\to\infty} &\tr(\bm{U}_o^T(\nabla\Capr(\bm{U}_t)-\nabla C_t(\bm{U}_t)))  \geq 0.  
\label{eq:thm1} 
 \end{align}
 Since $\bm{U}_o$ is arbitrarily chosen, \eqref{eq:thm1} can be true iff $\lim_{t\to\infty} \nabla\Capr(\bm{U}_t)-\nabla C_t(\bm{U}_t)) =0$.
\qedsymbol

\section{EXPERIMENTS}
\subsection{SYNTHETIC NETFLIX DATASET}

\subsubsection{Generating the data}
We conduct several experiments to  confirm that our online algorithm exhibits
 better results when we utilize the Laplacian.
 In particular, we first generate a synthetic Netflix dataset, 
 similarly as in  \cite{KaBrBrVa14}; the matrix that springs
 from this dataset obeys both
 the low--rank as well as the graph structure properties.
 The rows of the matrix represent   users and the columns represent   movies;
 the corresponding entries denote the rating.
 We consider a number $m_c=10$ of communities, forming a partition for the rows. The underlying graph is constructed as follows: two individuals are adjacent in the graph if and only if they belong to the same community. Similarly we assume that we have $n_c=20$ communities for the columns. 
The data matrix is then constructed by assigning a random value from $\lbrace 1, \dots, 5 \rbrace$ to each couple (movies community, users community).

It can be readily seen that this ideal matrix is of rank $r=\min (m_c,n_c)$.
 However, in practice it is very unlikely that we are dealing with low rank matrices, since the movie ratings are not necessarily consistent inside one users' community. And neither do we observe the communities one after another, because the order of appearance of the individuals is randomized. Therefore, to get a more realistic situation, we add   noise and we also permute all the columns. 

\subsubsection{Generating the noise}
The process to generate the noise in the Netflix framework is the following. Assuming that an user is likely to have a different opinion on a movie than the rest of his community, we define $N_{prob} \in [0 \dots 1]$ the probability of a rating to be affected by the noise, and $N_{level} \in \lbrace 1 \dots 5 \rbrace$ the maximum level of noise. Then, for each entry $X_{ij}$ of the data matrix, we pick the parameter $a$ according to a Bernoulli $\mathcal{B}(1, N_{prob})$ distribution  and the parameter $b$ according to the uniform $\mathcal{U}(\lbrace -N_{level}, -N_{level}+1, \dots, N_{level}-1, N_{level} \rbrace)$ distribution. The entry of the corresponding corrupted matrix is then defined as:
$$\tilde{X}_{ij} = \max(\min(X_{ij} + ab, 5), 1)$$
One can easily verify that this definition preserves the fact that the occurring noisy entry will belong to the
 $\lbrace 1, \dots, 5 \rbrace$ set.

\subsubsection{Error measurement}
We run the online algorithm and compute for each time step the euclidean distance between the predicted vector, $\hat{\bm{x}}_i = \bm{U}_i\bm{r}_i$, and the true one, divided by the norm of the latter. 	Afterwards, we compute the mean over time and the resulting metric is given by:
$  err(t) =20\log_{10}\left( \frac{1}{t} \sum_{i=1}^t \frac{\left\Vert \hat{\bm{x}}_i - \bm{x}_i\right\Vert_2}{\left\Vert \bm{x}_i \right\Vert_2}\right).  $

\subsubsection{Results}

In the following we study the realistic case of 20\% missing entries in the observations. We assume the sampling of the entries is uniform, which may not be true in practice. 
However the case of non-uniform sampling goes beyond the scope of this paper.
 
\begin{figure}[h]
\centering
\includegraphics[width=0.7\columnwidth]{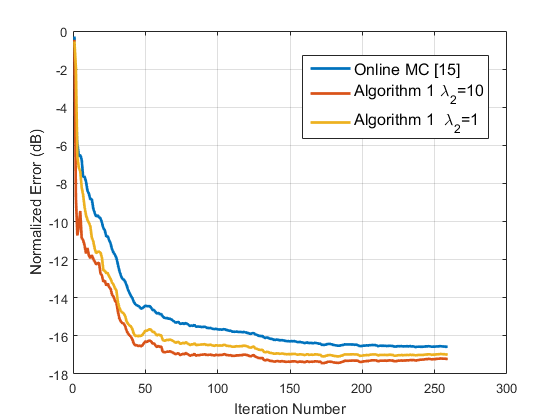}
\caption{Errors for Netflix dataset}
\label{fig:parameters}
\end{figure}

We compare our proposed algorithm with the one presented in \cite{giannakis}. 
This methodology is suitable for robust online matrix completion,
albeit no graph information is included. It is worth pointing out that, in both algorithms
the regularization parameter related to the sparse outlier 
noise is set equal to zero, since in this experiment we do not assume outliers. 
The rest parameters are chosen via cross validation
so that all the algorithms exhibit the best trade--off between convergence speed
and steady state error floor.
Contenting ourselves with a small level of noise ($N_{prob} = 0.3$ and $N_{level} = 1$), we obtain
 the results presented in Figure~\ref{fig:parameters}. 
It can be readily observed that the
Laplacian regularization improves the  performance, as expected. 
In fact, Algorithm 1 converges faster to a lower steady state error floor,
compared to the algorithm in \cite{giannakis}. Moreover, setting $\lambda_2=10$ exhibits a slightly improved performance compared to the $\lambda_2=1$ case.

%
%

\subsection{COTINUOUS VALUES DATASET: THE ROBUST CASE}
\begin{figure}[h]
\centering
\includegraphics[width=0.7\columnwidth]{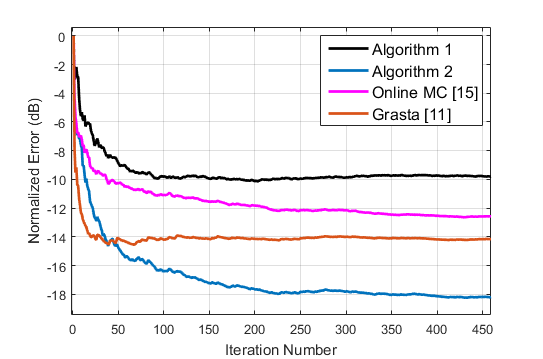}
\caption{Comparison of the standard and robust methods. Continuous values.}
\label{fig:robust3}
\end{figure}

In the previous experiment,  the entries of the data matrix are integers taking values between 1 and 5.
Such experiments do not permit us to evaluate if the robust algorithm deals well with 
very large (outlier) values.
 Therefore, we turn our focus now on another dataset generated in a similar way   
 as in the previous experiment,
 albeit the entries now are allowed to take continuous values.  To that end, they  are drawn from a zero--mean normal distribution with variance equal to $1$. We add i.i.d. Gaussian noise, with standard deviation equal to
$\sigma = 0.2$. On top of that, we add an ``outlier" sparse matrix, the non-zero entries of which have a high magnitude compared to the data matrix. The sparse matrix is generated randomly  and $1\%$ of its entries are non-zero.
 These non-zeros entries are constructed so that their magnitude is at least
  10 times the maximum value of the data matrix. Doing so, we have significant outliers.
We compare the proposed robust algorithm (Algorithm 2)  with: 
a) the non--robust one (Algorithm 1), b) a grassmannian manifold based algorithm
suitable for online robust matrix completion, \cite{Grasta} and c) 
the algorithm of \cite{giannakis}. Again, the parameters are chosen via cross validation.
 Figure~\ref{fig:robust3} presents the evolution of the error at each time step. 
It can be readily seen that, Algorithm 1 converges to a high error floor,
since the presence of outliers is not taken into account.
Furthermore, the proposed algorithm outperforms
the other robust based schemes, since it exploits the underlying
graph structure. 

\subsection{REAL NETWORK DATA}
\begin{figure}[h]
\centering
\includegraphics[width=0.7\columnwidth]{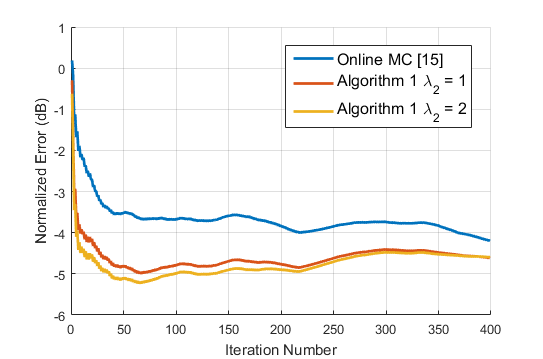}
\caption{Comparison of the proposed algorithm using the GEANT database}
\label{fig:robustnet}
\end{figure}

Let us now evaluate our proposed algorithm using data collected from a real network.
In particular, we use  the dataset captured in 2006,~\cite{Uhlig06} on GEANT, the high bandwidth pan-European research and education backbone. The network comprises $22$ nodes and $36$  links. 
We consider that at each time step, the load from a subset of the links
becomes available to us, whereas the load for the rest of them is unknown.
 Our goal is to estimate the load for these links.
 To that direction, we employ the proposed algorithm (Algorithm 1),
  for different values of the
Laplacian related regularization parameter $\lambda_2$,  as well as 
the online matrix completion algorithm of \cite{giannakis}.
In all the algorithms, we fix $\lambda_1$ to be equal to $0.1$,
since this particular choice leads to a fast convergence speed
 and a low steady state error floor at the same time.
  Moreover, in both algorithms
 the regularization parameter associated to the sparse outlier term
 is set equal to zero, since in that context there are no outliers.
 The results are shown in Fig.~\ref{fig:robustnet}.
 First, it should be highlighted that the online matrix completion algorithm 
 is able to provide a decent estimate of the missing entries due to the
 low--rank property of the link load traffic matrix.
 To be more specific, the network traffic pattern is highly correlated both temporally and spatially (i.e., across different links). This amounts to claiming that the data exhibit a 
 low rank structure.
  Nevertheless, the results can be enhanced significantly if we  
   exploiting the network graph topology, via the Laplacian smoothing.
  
\newpage

\bibliographystyle{plain}
\bibliography{refs}

\end{document}